\documentclass[useAMS,usenatbib]{mn2e}
\usepackage{amsmath}
\usepackage{graphicx}

\newcommand{\bt}{\mbox{\boldmath{$\theta$}}}

\newcommand{\drm}{\mathrm{d}}
\newcommand{\im}{\mathrm{i}}

\def\ave#1{\left\langle #1 \right\rangle}

\def\part#1#2{{\partial #1\over\partial #2}}

\def\A{{\cal A}}

\def\D{{\cal D}}

\def\P{{\cal P}}

\begin{document}
\label{firstpage}
\title{Probing the cosmic web: inter-cluster filament detection using gravitational lensing } \author[J. M. G. Mead, L. J. King and I. G. McCarthy] {James M. G. Mead$^{1, 2}$\thanks{Email: jmead@ast.cam.ac.uk}, Lindsay
  J. King$^{1, 2}$\thanks{Email:ljk@ast.cam.ac.uk} and Ian G. McCarthy$^{2, 1, 3}$\vspace{0.3cm}\\$^1$Institute of Astronomy, University of Cambridge, Madingley Rd, Cambridge CB3 0HA\\
$^2$Kavli Institute for Cosmology, University of Cambridge, Madingley Road, Cambridge CB3 0HA\\
$^3$Astrophysics Group, Cavendish Laboratory, Madingley Road, Cambridge CB3 0HE}
\date{}
\maketitle
\pagerange{\pageref{firstpage}--\pageref{lastpage}} \pubyear{2009}

\begin{abstract} 
The problem of detecting dark matter filaments in the cosmic web is considered. Weak lensing is an ideal probe of dark matter, and therefore forms the basis of particularly promising detection methods. 
We consider and develop a number of weak lensing techniques that could be used to detect filaments in individual or stacked cluster fields, and apply them to synthetic lensing data sets in the fields of clusters from the Millennium Simulation. These techniques are multipole moments of the shear and convergence, mass reconstruction, and parameterized fits to filament mass profiles using a Markov Chain Monte Carlo approach. In particular, two new filament detection techniques are explored (multipole shear filters and Markov Chain Monte Carlo mass profile fits), and we outline the quality of data required to be able to identify and quantify filament profiles. We also consider the effects of large scale structure on filament detection. We conclude that using these techniques, there will be realistic prospects of detecting filaments in data from future space-based missions.  The methods presented in this paper will be of great use in the identification of dark matter filaments in future surveys.
\end{abstract}


\section{Introduction}
In a universe dominated by Cold Dark Matter (CDM) such as our concordance cosmology, $\Lambda$CDM, hierarchical structure formation takes place with the growth of collapsed objects progressing via merging of smaller objects. Simulations of structure formation show that galaxy clusters are located at the intersections of filaments, which form a web-like structure throughout the universe. The processes of merging and continuous accretion of surrounding matter are thought to occur highly anisotropically,  with the filaments channeling matter along preferred directions. Modern galaxy surveys such as 2dFGRS (e.g. \cite{Coll01}) and SDSS (e.g. \cite{yorksdss}, \cite{Adel08}) have provided a dramatic picture of this so-called `cosmic web', where voids lie between filamentary arms traced by galaxies. From a theoretical perspective, \citet{Bond96} developed a theory for the cosmic web in a CDM cosmogony, where filaments are a product of a primordial tidal field evolving under non-linear effects.

Detection of inter-cluster filaments has important implications for cosmology, confirming the prediction that structure grows highly anisotropically from small initial perturbations in the cosmic density field. Their detection provides an important validation of our picture of hierarchical structure formation and cluster evolution \citep{Colfil}.  Dark matter filaments also have a significant role to play when considering the total mass budget of the universe - the galaxies and gas channeled along these filaments may account for as much as half of the baryonic mass budget in the universe. 

Inter-cluster filaments have been the target of a number of observational searches. The earliest of these used bremsstrahlung X-ray emission to detect hot gas being channeled along filaments. For example, \cite{Briel95} used X-ray data from the ROSAT All-Sky Survey, but failed to find evidence for filaments. Subsequent X-ray searches have also proved inconclusive (\cite{scharf00}, \cite{KB99}, \cite{durret03}), the main issue being that it is difficult to ascertain whether strong X-ray emission is due to matter in a filament or due to the clusters interacting. Another proposed observational method looks for overdensities of galaxies relative to the background. \cite{PD04} reported finding a short filament between galaxy clusters A1079/1084 using such a method. \cite{Ebeling} report the detection of a filament leading into MACS J0717.5+3745. However, this approach is susceptible to large errors and requires spectroscopic data to establish the redshifts of the target galaxies. An alternative detection technique uses a `skeleton', which is defined as the subset of critical lines joining the saddle points of a field to its maxima while following the gradient's direction. The skeleton formalism has been used in numerical simulations to extract and analyze the filamentary structure of the universe (\cite{Nov}, \cite{sousbie}). 

Work by \cite{Hahn} and \cite{Aragon} demonstrates that it is possible to use the Hessian of either the potential or density field to provide a dynamical classification of filamentary structure. This work was extended by \cite{romero}. The number of eigenvalues with magnitude above a certain threshold at each grid point determines whether that point belongs to a void, sheet or filament. \cite{Hahn} investigated the effect filaments have on the properties of dark matter halos, and found that filaments influence the magnitude and direction of halo spin as well as halo shape. There are a number of other filament detection methods, not elaborated here, which are discussed in \cite{Pim05}.

One of the most promising methods of filament detection is weak gravitational lensing. Since gravitational lensing is insensitive to whether the matter is luminous or dark, and to its dynamical state, it is an ideal probe of dark matter. A number of attempts have been made to image filaments using weak lensing. \citet{Gray02} claimed to detect a filament in the A901/902 supercluster, but this was later shown by \cite{Heymans08} to be an artifact of the mass reconstruction. \cite{Kaiser98} reported a detection of a filament between two of the galaxy clusters in the supercluster MS 0302+17, but these results could not be reproduced by \cite{gav04}, and hence this detection is called into question. \cite{Dietrich} used a variety of methods to detect a filament between A222/223 - however, they point out that it is difficult to objectively define whether this lensing signal is caused by a real filament. The Hubble Space Telescope Cosmic Evolution survey (COSMOS) has provided corroboratory evidence for a cosmic network of filaments \citep{Massey}.

Although there have been a number of observational attempts to detect the filamentary cosmic web with various degrees of success, conclusive evidence has proved elusive in all but a few cases, and widely applicable methods are still lacking. In this paper we assess existing techniques to detect filaments in weak lensing data and present two new, more effective, detection methods. In particular we present a method based on multipole filtering of the shear field, and a method based on Markov Chain Monte Carlo techniques that allows both detection and quantification of filament profiles. Motivated by the promise of weak lensing data from many thousands of galaxy clusters from large surveys such as the Dark Energy Survey (DES) and Large Synoptic Survey Telescope (LSST) we also consider detection in stacked data sets from a number of clusters. These methods are then applied to single and stacked synthetic weak lensing data sets from the fields of clusters from the Millennium Simulation \citep{springel05}. We will show that weak lensing is an ideal and versatile tool to detect dark matter filaments.

This paper is organized as follows. After a review of the basic lensing formalism (section 2), in section 3 we discuss the various methods by which we may detect filaments using weak lensing. These methods are then applied to data sets from the fields of simulated clusters. In section 4 we describe these simulations and in section 5 we detail our results. In section 6 we analyze the effects of large scale structure noise on our results. Finally, in section 7 we present our discussions and conclusions.

\section{Summary of Weak Lensing Formalism}

In this section we present the basic formalism of weak gravitational lensing. The surface mass density of a lens at position $\bt$ is denoted by $\Sigma \left(\bt \right)$.  The critical surface mass density of a lens at redshift $z_d$, with background sources at redshift $z$, is defined as $\Sigma_{\rm{crit}} = \frac{c^{2}}{4 \pi G} \frac{D_{s}}{D_{d} D_{s}}$, where $D_{s}$, $D_{d}$, $D_{ds}$ are the observer-source, observer-lens and lens-source angular diameter distances respectively. The dimensionless surface mass density of the lens is called the convergence, and is defined by $\kappa \left( \bt, z \right) = \frac{\Sigma}{\Sigma_{\rm{crit}}}$.  The convergence is related to the deflection potential $\psi \left(\bf \theta \right)$ through a Poisson equation,
\begin{equation}
\nabla^{2} \psi = 2 \kappa\;.
\end{equation}
Whereas the convergence magnifies background objects such as galaxies, the shear stretches them tangentially. The complex shear depends on the second derivatives of the lensing potential,
\begin{equation}
\gamma = \gamma_{1} + i \gamma_{2} = \left(\psi_{,11} - \psi_{,22} \right)/2 + i \psi_{,12}\;,
\end{equation}
where commas denote partial differentiation.

The complex reduced shear is obtained from the shear and convergence through the equation $g=\gamma/(1-\kappa)$. The magnification is given by the inverse Jacobian determinant of the lens equation, $\mu(\bt) = [\det \A \left( \bt \right)]^{-1}$. Evaluating this we obtain,
\begin{equation}
\mu = \frac{1}{\left[\left(1-\kappa \right)^{2} - \gamma^{2} \right]}\;.
\end{equation}
We can define the ellipticity of a galaxy, describing its shape and orientation, by a complex number with a modulus related to the axis ratio $r$ of $(1-r)/(1+r)$, and phase being twice the position angle.
In the non-critical regime ($\det \A > 0$) the weakly lensed image of a distant galaxy has a complex ellipticity given by,
\begin{equation}
\epsilon=\frac{\epsilon^{\rm s}+g}{1-g^{*}\epsilon^{\rm s}}\;,
\end{equation}
\noindent where $\epsilon$ and $\epsilon^{\rm s}$ are the lensed and unlensed complex ellipticities respectively, and $g^{*}$ is the complex conjugate of the reduced shear. Galaxy ellipticities are taken to have a Gaussian probability density function, 
\begin{equation}
\label{epdf}
p_{\epsilon^{\rm s}}= \frac{\exp\left({-|\epsilon^{\rm s}|^{2}/\sigma_{\epsilon^{\rm s}}^{2}}\right)}{\pi\sigma^{2}_{\epsilon^{\rm s}}\left[1-\exp\left({-1/\sigma_{\epsilon^{\rm s}}^2}\right)\right]}\;.
\end{equation}
Throughout this work we take $\sigma_{\epsilon} = 0.2$. In weak lensing, since the distortion of any individual background galaxy is too small to detect, we must look at a large sample of galaxies and statistically detect the presence of dark matter, which will manifest itself in, for example, a non-zero mean ellipticity of background galaxies.  It has been shown that the expectation of the lensed ellipticity is $\langle \epsilon \rangle = g$ in the non-critical regime \citep{sk95} and  $\langle \epsilon \rangle = \frac{1}{g^{*}}$ in the critical regime \citep{seitsch}.

\section{Methods to Detect Filaments}

In this section we outline the methods of filament detection used in this paper: non-parametric mass-reconstruction, parameterized model fits using MCMC techniques, and aperture multipole moments. 

\subsection{Non-parametric Mass Reconstruction}
We perform finite-field non-parametric mass reconstructions using the algorithm of \cite{seitzrec}. Non-parametric mass reconstructions use a statistical analysis of many weakly lensed galaxies to identify over-densities in the mass distribution.  

Many previous attempts at detecting filaments using gravitational lensing have focused on cluster pairs, and in this paper we too will consider such filaments. This is because filaments between closely separated clusters tend to be both strong and regular. However, mass reconstructions must be applied with care to closely separated pairs of clusters. The smoothing length of the mass reconstruction represents the typical scale over which the resulting reconstructed mass is `smeared'. If this smoothing scale is greater than the inter-cluster separation, this smearing will join the two clusters, resulting in a possible false identification of a filament. Indeed, this was suspected to have been a factor in the false detection of a filament in the A901/902 supercluster (\cite{Gray02}; \cite{Heymans08}). 

\subsection{Parameterized Mass Models}

The probability distribution for the \emph{observed} galaxy ellipticities, $p_{\epsilon}$, can be obtained from the \emph{intrinisc} ellipticity distribution $p_{\epsilon^{\rm s}}$ (\ref{epdf}) as follows,
\begin{equation}
p_{\epsilon}(\epsilon |g) = p_{\epsilon^{\rm s}} \left(\epsilon^{\rm s} \left( \epsilon |g\right)\right)\left
|{\partial^{2}\epsilon^{\rm s}}\over{\partial^{2}\epsilon}\right | = p_{\epsilon^{\rm s}}(\epsilon^{\rm s}(\epsilon |g))
{(\left | g\right
|^{2}-1)^{2} \over\left |\epsilon g^{*}-1\right |^{4}}\;.
\end{equation}
Using this, we can compute the log-likelihood function from the probability density for each lensed galaxy, $p_{\epsilon}\left(\epsilon_{i}\right)$,
\begin{equation}
\ell_\gamma=-\sum_{i=1}^{N_\gamma}{\ln p_\epsilon(\epsilon_i|g(\bt_i))}\;,
\end{equation}
which can be evaluated numerically for a trial mass model (with corresponding reduced shear field $g(\bt)$) given the lensed ellipticities.

For a mass model with a small number of parameters to obtain the best-fit parameters it is often appropriate to use a minimization technique such as the conjugate gradient algortithm of \cite{Powell64}. In this paper we instead employ an MCMC approach, implementing our likelihood calculation as a module in {\small COSMOMC} \citep{lewbri}, using it as a generic sampler 
to explore parameter space with the standard Metropolis-Hastings algorithm (\cite{Met}; \cite{Has}). 

\subsection{Aperture Multipole Moments}

\cite{SB97} introduced the concept of an aperture multipole moment, defined by,

\begin{equation}
Q^{\left(n\right)}\left(\bt_{0} \right) = \int_0^\infty \drm^2 \bt\; \theta^{n} U \left(|\bt| \right) \mathrm{e}^{n \im \phi} \kappa \left(\bt_{0} + \bt \right)\;,
\end{equation} 

\noindent where $U \left(| \bt | \right)$ is a radially symmetric weight function, and we are using polar coordinates $\left(\theta, \phi \right)$. In particular, the aperture quadrupole moment of the convergence has been used by \cite{Dietrich} as a method for detecting filaments.  We can express the aperture moments in terms of shear estimates \citep{SB97},

\begin{eqnarray}
  \label{}
  &&Q^{(n)}\left( \bt_0 \right) = 
  \frac{1}{\overline{n}} 
  \sum_{i=1}^{N} \rm{e}^{n \im \phi_{\it i}} \times \nonumber\\
  && \left\{
    \theta_i^n U(\theta_i) \varepsilon_{\mathrm{t}i} +
    \im \frac{\theta_i^n [
      nU(\theta_i) + \theta_i U'(\theta_i)
      ]}{n} \varepsilon_{\times i}
    \right\}\;,
\end{eqnarray}

\noindent where $\overline{n}$ is the number density of galaxies in the aperture, and $\epsilon_{{\rm t}i} = - \mathcal{R}(\epsilon_i \rm{e}^{-2 i \phi_i})$ and $\epsilon_{{\times}i} = -\mathcal{I}(\epsilon_i \rm{e}^{-2 \rm{i} \phi_i})$ are the tangential and cross components of the shear respectively. $U'(\theta)$ is the derivative of the weight function with respect to the radial coordinate. \cite{Dietrich} used a weight function of the form,
\begin{equation}
U\left(\theta \right)=\left\{\begin{aligned}
   1 -& \left(\frac{\theta}{\theta_{max}} \right)^{2}\qquad \textrm{for }
\theta\leq\theta_{max}\\
    0 &\qquad \textrm{~~~~~~~~~~~~~~otherwise}
     \end{aligned}\;,\right.
\label{flimits}
\end{equation}
\noindent where $\theta_{\rm max}$ is the aperture radius. This serves well as a general multi-purpose filter, but may not be optimal for filament detection, as it does not match their expected mass profiles \citep{Colfil}.

In reality, a single order of multipole will not completely characterize a mass distribution - any given mass distribution will in general give non-zero results for a variety of different multipole orders. Therefore, instead of applying a quadrupole filter alone to detect a filament, a more optimal method would use  a superposition of filters thus,

\begin{equation}
\chi = Q^{(0)} + \alpha_{1} Q^{(1)} + \alpha_{2} Q^{(2)} + ... \;,
\label{superpose}
\end{equation}
\noindent where the constants $\alpha_{i}$ can be chosen to maximize the signal-to-noise.

We can also define multipole moments of the shear field,

\begin{equation}
\zeta^{(n)} \left(\bt_{0} \right) = \int_0^\infty \drm^2 \bt\; \theta^n U \left(| \bt | \right) e^{\im n \phi} \gamma \left(\bt_{0} + \bt \right)\;.
\label{shearint}
\end{equation}

This can be related to the convergence through the relation,

\begin{equation}
\gamma \left( \bt \right) = \frac{1}{\pi} \int \drm^2 \bt\; \theta'  \D(\bt - \bt') \kappa(\bt)\;,
\end{equation}

\noindent where,

\begin{equation}
\D(\bt) = \frac{\theta_1^2 - \theta_2^2 + 2 \im \theta_{1} \theta_{2}}{|\bt|^{4}}\;.
\end{equation}

The integral eq.\,(\ref{shearint}) can also be written in terms of shear estimates to give,
\begin{eqnarray}
  \label{shearfilt}
  \zeta^{(n)}\left( \bt_0 \right) & = & 
  \frac{1}{\overline{n}} 
  \sum_{i=1}^{N} e^{\left(n+2\right) i \phi_i} \theta_{i}^{n} U(\theta_i) 
	\left[\epsilon_t +  \im \epsilon_\times \right]\;.
\end{eqnarray}

To calculate the noise associated with either the shear or convergence multipole filters, we use the result of \cite{SB97}. Both the shear and multipole filters have the form,
\begin{eqnarray}
&&\eta^{(n)} \left\{\bt_{0} \right) = \int \drm^2 \bt \; \rm{e}^{n \im \phi}\times\nonumber\\
&&\left\{\frac{{\it b}_{\rm{t}}(\theta)}{\theta} \gamma_t (\bt; \bt_0)+ \im \frac{{\it b}_\times (\theta)}{\theta} \gamma_\times (\bt; \bt_0) \right\}\;,
\end{eqnarray}
\noindent where $b_{\rm{t}} (\theta)$ and $b_{\times} ( \theta)$ are functions of the radial distance coordinate and $\gamma_{\rm t}$ and $\gamma_{\times}$ are defined by the following,
\begin{equation}
\gamma_{\rm t}\left(\bt;\bt_{0}\right)=-{\mathcal R}\left[\gamma\left(\bt+\bt_{0}\right){\rm e}^{-2\im\phi}\right]\;,
\end{equation}
and 
\begin{equation}
\gamma_{\times}\left(\bt;\bt_{0}\right)=-{\mathcal I}\left[\gamma\left(\bt+\bt_{0}\right){\rm e}^{-2\im\phi}\right]\;.
\end{equation}

Using the result of \cite{SB97} for filters of this form, the corresponding ensemble-averaged noise is,
\begin{eqnarray}
  \label{}
  \left[ \sigma_c^{(n)} \right]^2 = \frac{\pi \sigma_\epsilon^2}{\overline{n}}
  \int_0^{\theta_{\rm{max}}} \drm y \; \left[\frac{b_{\rm{t}}^2 (y) + b_{\times}^2 (y)}{y} \right]\;.
\end{eqnarray}

In this paper, we define the signal-to-noise ratio of a multipole filter by,

\begin{equation}
  \label{}
  S^{(n)} = \frac{| \eta^{(n)} |}{\sigma_c^{(n)}}\;.
\end{equation}

\subsection{Stacking}

Since filaments are very weak, their detection will usually require high background galaxy counts, above that typically achievable at present. One solution to this problem would be to stack catalogues of background galaxies in the cluster fields. Once these lensed catalogues have been stacked, one can apply the techniques described above to search for structures of interest. 

In this paper, we use stacked catalogues to increase the ease with which we are able to detect filaments. For stacking to improve our chances of detection, we need to align filaments in the stacked data. Since we use stacked catalogues to detect filaments that cannot be detected on single cluster images, we must infer the probable filament direction before stacking the data. There are two possible ways of doing this. The first method uses the result that a filament is often aligned with the major axis of a cluster \citep{Colfil}. Since we can identify the cluster major axis from a mass reconstruction, we can then align these major axes in our stacked catalogues. Depending on the quality of the mass reconstruction or parameterized model fit, one could also align cluster centres using optical images (assuming that light traces mass). The second method involves using cluster pairs as `markers' for the starts and ends of filaments. It is expected that cluster pairs will be connected by filaments, so by aligning the axis running through the cluster centres we also align any filament between them. The cluster pairs must be close enough to ensure that a filament connects them, but separated by several virial radii to minimize contamination by a cluster signal. 

Both methods have advantages and disadvantages, and the most useful method will depend on the data-set being analyzed. The first method can be applied to all clusters in a sample, whereas the second requires cluster pairs with a particular spacing, which are rarer. However, the second method has the advantage that filaments between cluster pairs tend to be straighter and more regular than those filaments which do not connect to another cluster \citep{Colfil}. If we simply stack the major axes, a significant fraction of the filaments will be very diffuse, `warped' or irregular - stacking such filaments does not increase the ease of detection, as there will be little overlap of irregular filaments from each cluster. Clusters with similar separations could also be stacked to maximize the extent of filament available for analysis. In this way, the dependence of filament properties on inter-cluster separation can also be assessed.
\begin{figure*}
\centering
     \begin{tabular}{cc}
\includegraphics[width=3.5in]{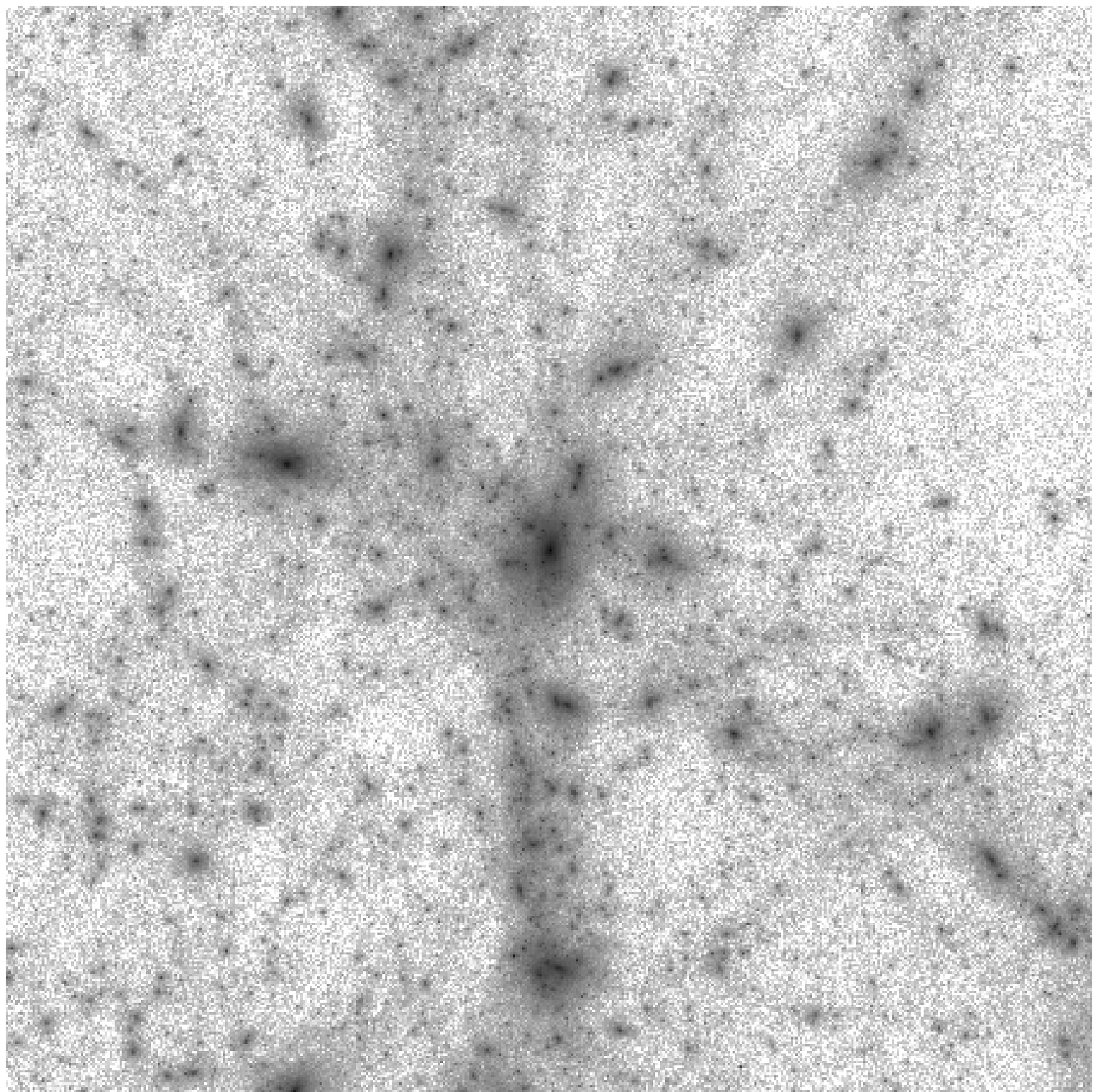}&
\includegraphics[width=3.5in]{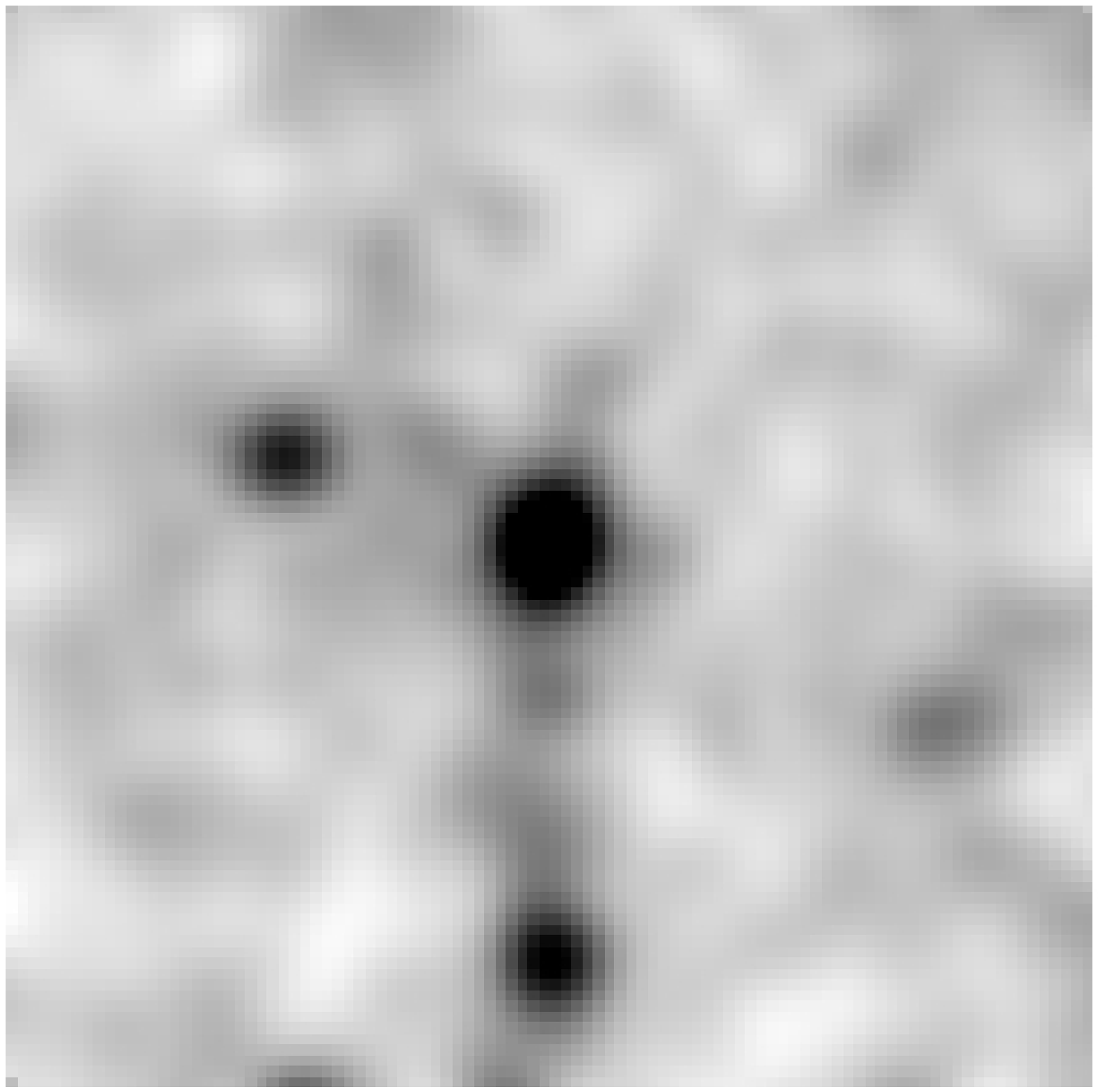}\\\\
\includegraphics[width=3.5in]{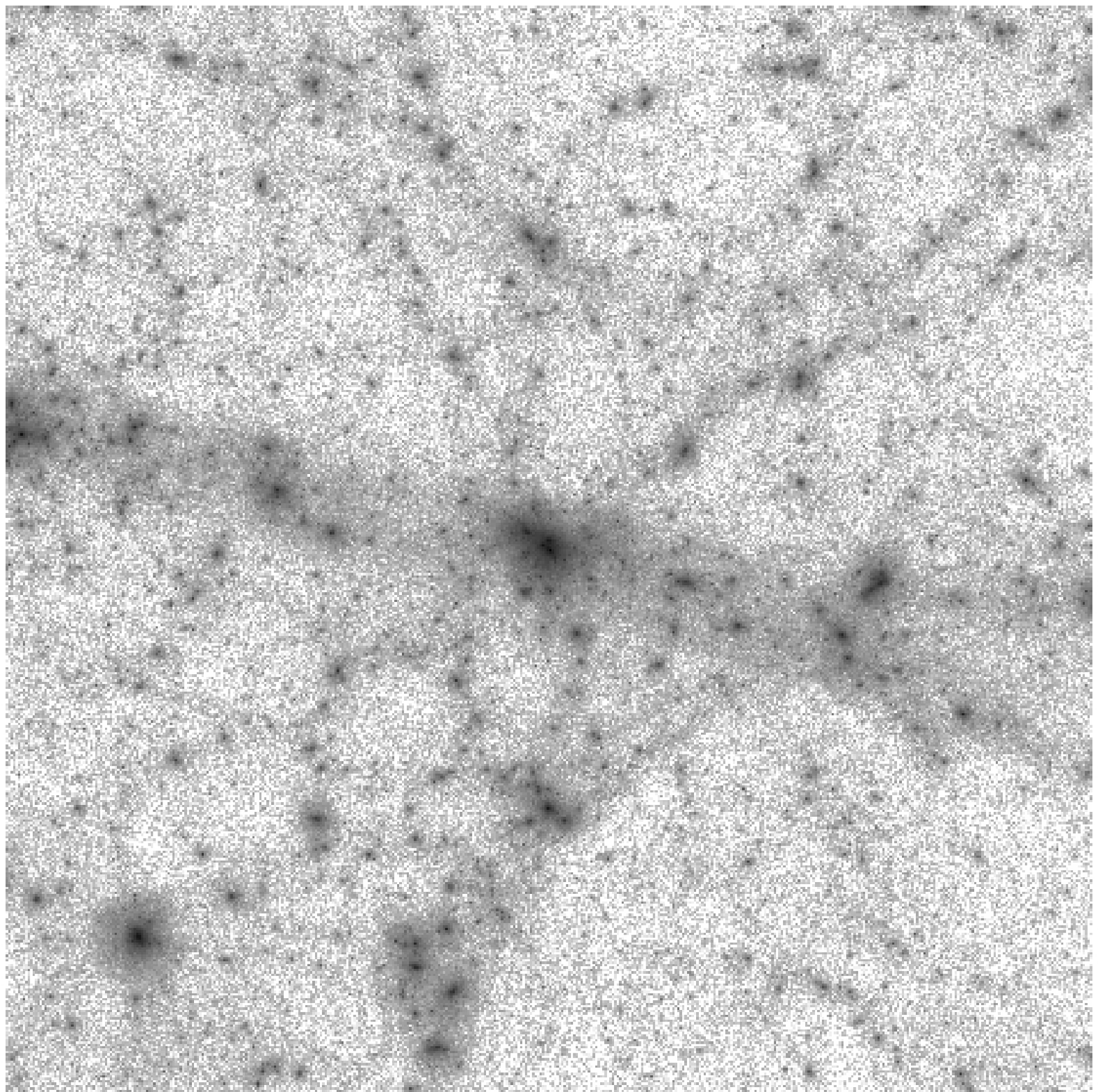}&
\includegraphics[width=3.5in]{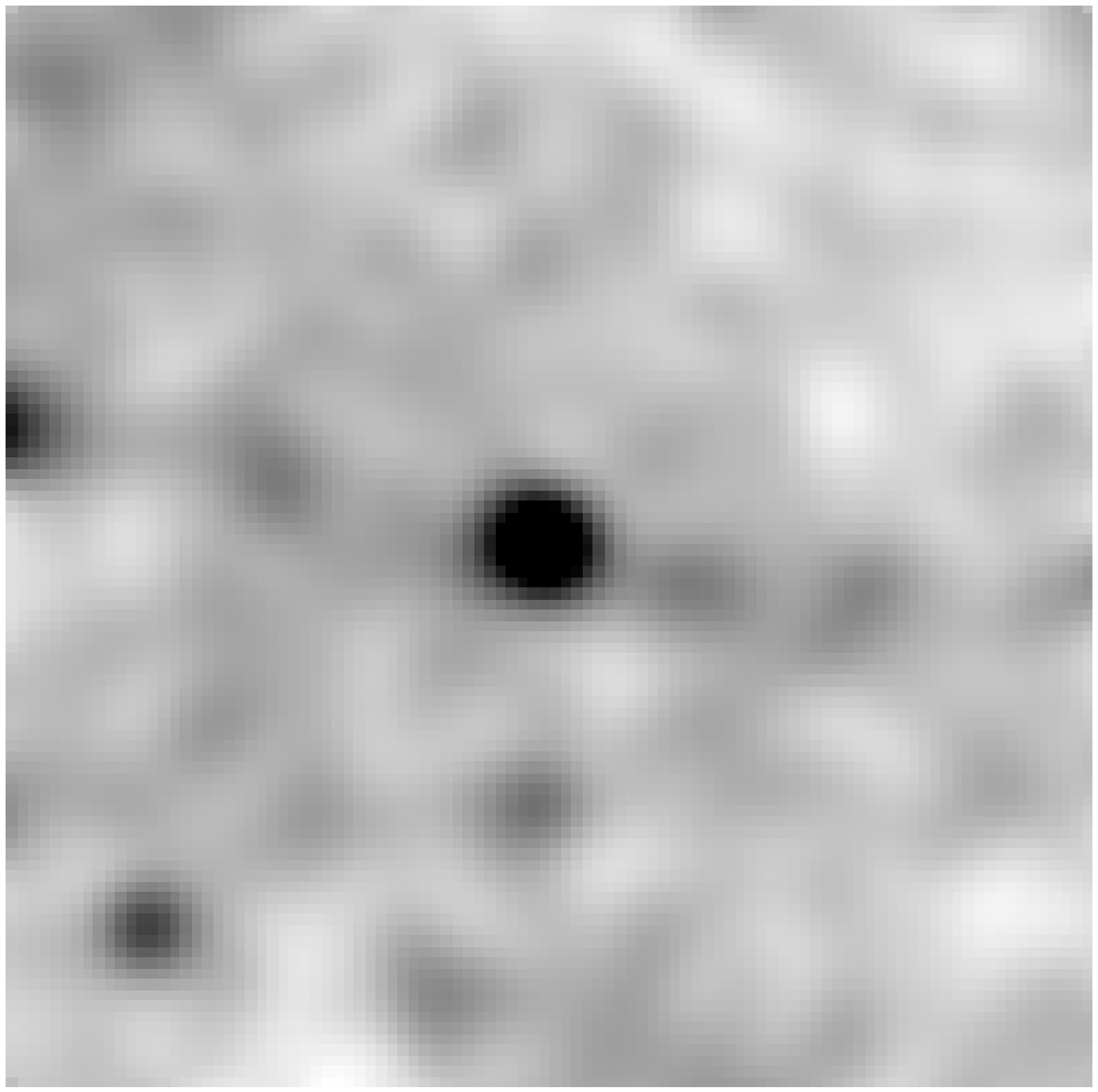}\\\\
 \end{tabular}
\caption{In this figure, the left-hand panels show the simulated cluster convergence maps on a logarithmic scale, whereas the right-hand panels show the corresponding mass reconstructions. The background galaxy density is 100 arcmin$^{-2}$, and the smoothing scale is 1.5\,$h^{-1}$\,Mpc. The side length of the field is 3.5$^{\circ}$, and the reconstruction is calculated on a 100 $\times$ 100 pixel grid. We see that the reconstruction provides some evidence for strong filaments, but is unable to detect weaker filaments.}
\label{singlemassrec}
\end{figure*}

\section{Numerical Simulations: Clusters and Sources }
For this study we make use of the Millennium Simulation \citep{springel05}, a large cosmological N-body simulation that follows $2160^3$ dark matter particles from $z=127$ to $z=0$ in a periodic box of $500\,h^{-1}$\,Mpc on a side.  The cosmology adopted for the Millennium Simulation (MS) is a flat $\Lambda$CDM model with $h=0.73$, $\Omega_M =0.25$, and a power spectrum normalisation on a scale of $8 h^{-1}$Mpc (i.e., the rms linear mass fluctuation in a sphere of radius $8 h^{-1}$ Mpc extrapolated to present-day) of $\sigma_8 = 0.9$. These parameters are consistent with the latest measurements of temperature and polarization anisotropies in the cosmic microwave background (Komatsu et al.\ 2009), although the value of $\sigma_8$ adopted for the MS is larger than the maximum-likelihood CMB value by $\sim 2\sigma$.  The larger value of $\sigma_8$ means that the MS will have more massive clusters than than a universe with the WMAP 5-year cosmology.  However, for the purposes of our study this is not important as we are interested in the lensing signal (and structure) of {\it individual} clusters and their filaments and not the statistics of cluster population as a whole.  The mass structure of individual clusters (e.g. halo concentration), and presumably their filaments, depends only weakly on cosmological parameters (e.g. Duffy et al.\ 2009) and therefore the exact choice of parameters is not expected to influence our results or conclusions.

From the MS, we select the 100 most massive galaxy clusters at $z_{\rm c}=0.2$ for
analysis.  These are the systems which are expected to have the most prominent filaments.  The clusters are identified using a standard friends-of-friends (FOF) algorithm with a linking length of 0.2 times  
the mean interparticle separation.  The clusters we have selected range in mass from $M_{200} = 3.90\times10^{14}\,h^{-1}\,M_\odot$ to $M_{200} = 2.16\times10^{15}\,h^{-1}\,M_\odot$ (for comparison, the mass resolution of the MS is $8.6\times10^8 h^{-1} M_\odot$.)  We extract all of the dark matter particles in a $50\,h^{-1}$\,Mpc box centred on the FOF halo centre of mass.  For each box, we generate surface mass density maps by projecting along the axes of the box; we make 3 separate projected maps, by projecting along the x-, y-, and z-axes of the box.  The dark matter particles are interpolated onto a regular 2D grid (map) using a triangular shaped clouds interpolation alogrithm.  We fixed the  
number of pixels in the 2D grids to be $4000^2$, implying a pixel length of $12.5\,h^{-1}$\,kpc, which is comparable to the gravitational softening length of the MS (which is $5\,h^{-1}$\,kpc).  As we shall see, this resolution is more than sufficient for making weak lensing predictions of clusters and their
filaments for current and future (e.g., space-based) observations; the number densities of sources imply  
effective resolution well in excess of several tens of kpc.

Simulated galaxy catalogues are created by assigning galaxies random positions in the field, and by simulating galaxy ellipticities as Gaussian random deviates. A magnification bias is also included in the simulated population (e.g. \cite{Can82}), which changes the galaxy number counts after weak lensing. The local cumulative number counts $n \left(\bt; S \right)$ above a flux limit $S$, are related to the unlensed counts $n_{0} \left(S \right)$ by,
\begin{equation}
n \left(\bt; S \right) = \frac{1}{\mu} n_{0} \left( \frac{S}{\mu} \right)\;, 
\end{equation}
where $\mu$ is the magnification. If we assume that the number counts locally follow a power law of the form $n_{0} \propto S^{- \beta}$, then,
\begin{equation}
n \left( \bt \right) = n_{0} \mu \left(\bt\right)^{\beta -1}\;,
\end{equation}
at any fixed threshold. This implies that if the intrinsic counts are flatter than 1, then the lensed counts will be reduced relative to the unlensed counts. In this paper we use $\beta = 0.5$, typical of the faint galaxy populations studied in weak lensing. To account for the effect of magnification we reject a given galaxy from the catalogue if a uniform random deviate in the range $[0,1]$ is greater than $\mu^{\beta - 1}$. This means that the higher the magnification of a given galaxy, the more likely it is to be excluded from the catalogue.

In our analysis, we assume that the filaments associated with a cluster are at the same redshift as the cluster itself. Thus our typical simulation box of 50\,$h^{-1}$\,Mpc is projected onto a lens plane at the cluster redshift, $z_{\rm c} = 0.2$. The cluster resdhift is sufficiently low such that the redshift distribution of the source galaxy population can be neglected (e.g. \cite{KS01}); we take the source galaxy redshift to be $z_{\rm g} = 1$, characteristic of fairly deep ground-based observations.

\section{Results}
We investigated several methods for filament detection: mass reconstructions, multipole filters in both the convergence and shear, as well as MCMC fits to filament profiles. In this section we describe the results of these techniques. In each case, we analyzed a square field centred on the target cluster, with a side-length of 3.5$^{\circ}$.

\subsection{Mass Reconstruction and Aperture Convergence Multipole Moments}
We performed finite field mass reconstructions on single galaxy cluster synthetic data sets using background galaxy densities of 30 arcmin$^{-2}$ and 100 arcmin$^{-2}$, the former being typical of current ground-based observations and the latter being representative of future space-based missions. The results are shown in Fig. 1. With a background galaxy density of 30 arcmin$^{-2}$, filaments are not detected in reconstructions. With a background galaxy density of 100 arcmin$^{-2}$,  strong filaments are detected more clearly, but weaker filaments remain undetected. These strong filaments are all characterized by having regions of the filament which contain large clumps of dark matter (corresponding to massive galaxy groups or poor clusters), and it is these regions that the mass reconstruction is able to detect.  In Fig.\,\ref{singlemassrec} there is visual evidence for particularly strong filaments in clusters, but not for weaker ones which are missed.

The multipole filter outlined in eq.\,(9) was applied to the same data sets, using $n=1$ and $n=2$ (a dipole moment and a quadrupole moment). A variety of filter functions and aperture sizes were tried, but no filaments were detected by either multipole moment. We also superposed the two multipole moments as per eq.\,(\ref{superpose}), but were unable to make a filament detection. We lowered the ellipticity dispersion in the synthetic galaxy catalogues to $\sigma_{\epsilon} = 0.05$, at which point both $Q^{(1)}$ and $Q^{(2)}$ registered weak detections of the strong filaments. Although this $\sigma_{\epsilon}$ is unrealisitic, it allows us to make a rough estimate of how many clusters we would need to stack for these multipole moments to become effective. Taking the noise to vary as $\frac{1}{\sqrt{N}}$, where $N$ is the number of stacked clusters, we estimate that we would need to stack approximately 20 clusters to detect filaments using this method. This is likely to be a lower bound, due to imperfect filament alignment and the presence of irregular filaments in the stacked clusters.

\begin{figure}
\centering
\includegraphics[width=.45\textwidth]{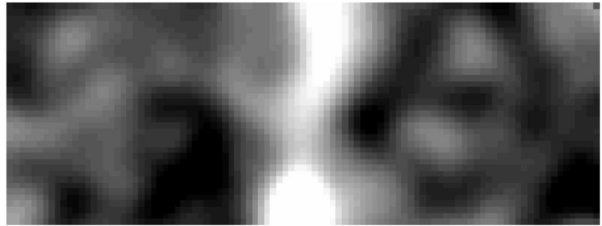}
\caption{Mass reconstruction for 10 stacked clusters with a background galaxy density of 30 arcmin$^{-2}$. Each stacked data set corresponds to a cluster pair, with each individual data set rotated such that the cluster-cluster axes are aligned in the stacked catalogue.  The smoothing scale used in the reconstruction was 1.5\,$h^{-1}$\,Mpc. The region shown above is the inter-cluster region, with a vertical height of 1.3$^{\circ}$ and a horizontal length of 3.5$^{\circ}$. We see clear evidence for filamentary presence between the cluster pairs.} 
\label{massrec}
\end{figure}

The mass reconstruction was then applied to stacked data fields using both of the stacking methods outlined in section 3.4. As predicted, stacking using double clusters proved to be more effective, and in Fig.\,\ref{massrec} we show the mass reconstruction for 10 stacked galaxy fields, rotated such that the axis joining the centres of the two clusters lies in the vertical direction. To select cluster pairs, we identified all those fields for which one cluster was observed to be within 25\,$h^{-1}$\,Mpc of another cluster. The density of background galaxies used for each cluster was 30 arcmin$^{-2}$, and the smoothing scale used was 1.5\,$h^{-1}$\,Mpc, chosen to approximately correspond to the predicted width of a filament (\cite{Colfil}). We see clear evidence of filamentary extension in the inter-cluster region. The region shown in Fig.\,\ref{massrec} lies outside the virial radii of both clusters - in other words, clusters don't contribute to the signal. A mass reconstruction could also detect evidence for filaments when major axes of clusters were stacked in the vertical direction, but this signal was weaker than that shown in Fig.\,\ref{massrec}.

We applied the first and second multipole moments $Q^{(1)}$ and $Q^{(2)}$ to the stacked data. We detected a faint $Q^{(1)}$ signal and no $Q^{(2)}$ signal.

\subsection{Shear Multipole Moments}

\begin{figure*}
\centering
     \begin{tabular}{cc}
\includegraphics[width=3.5in]{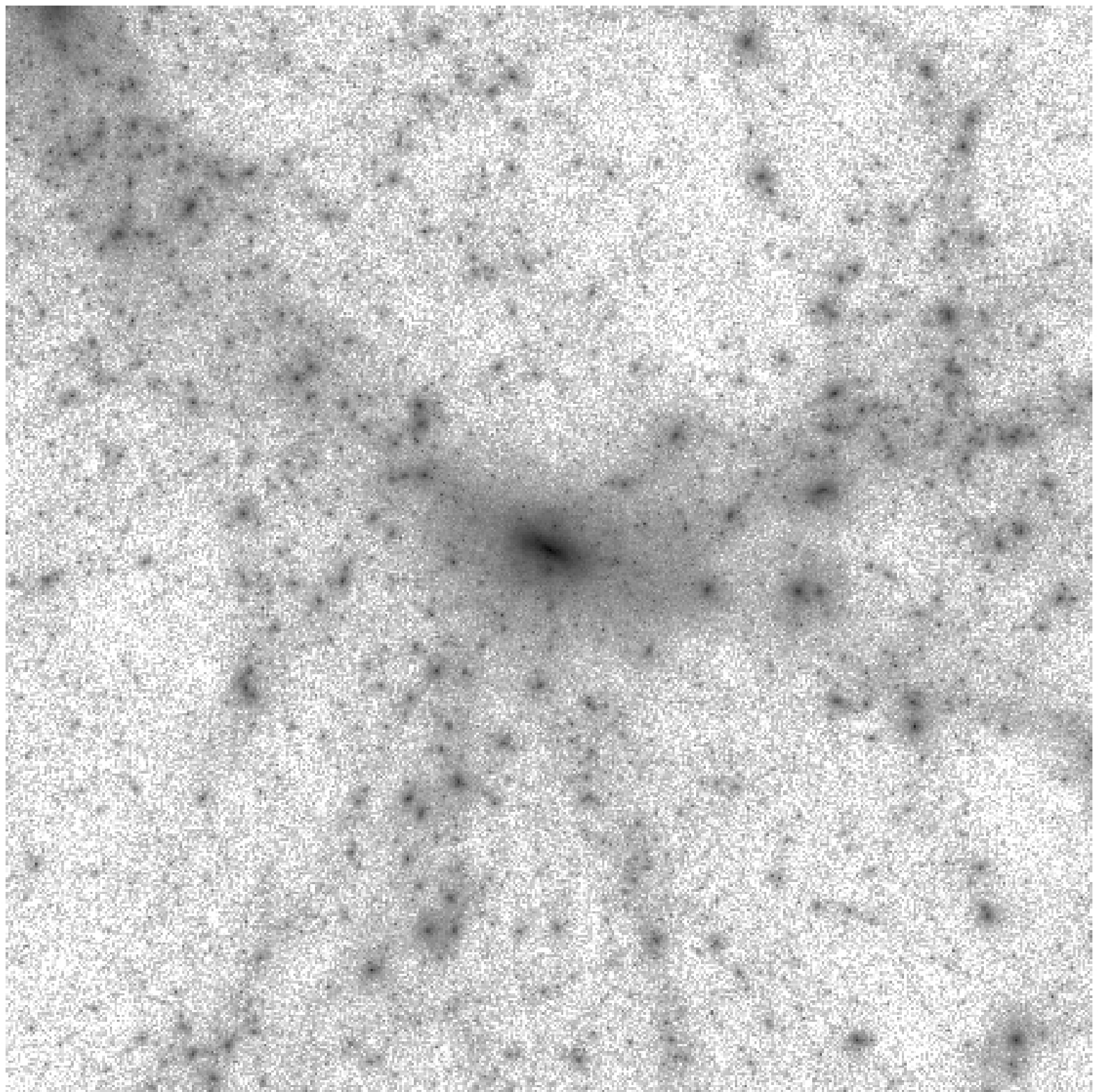}&
\includegraphics[width=3.5in]{cluster48.eps}\\\\
\includegraphics[width=3.5in]{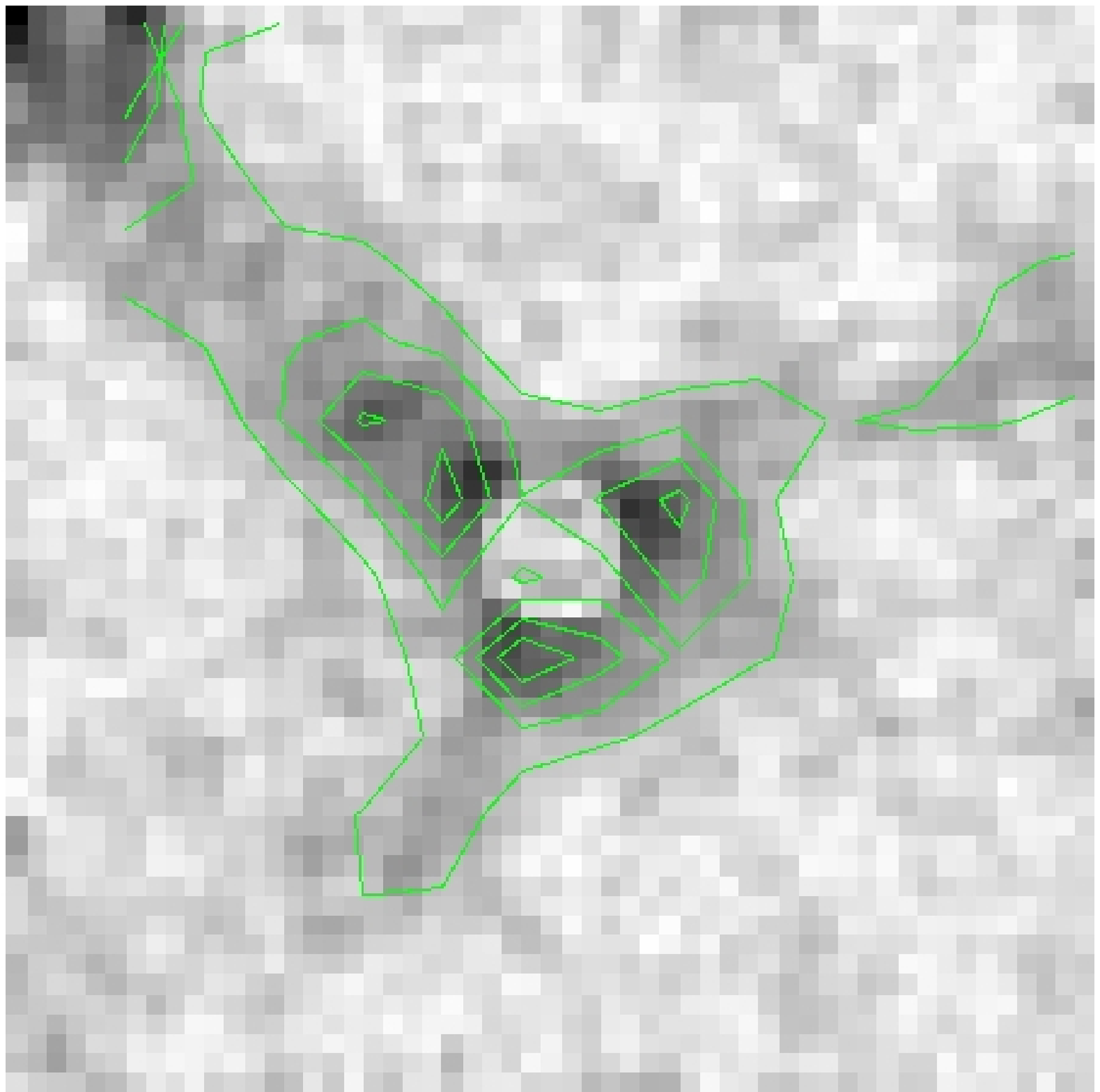}&
\includegraphics[width=3.5in]{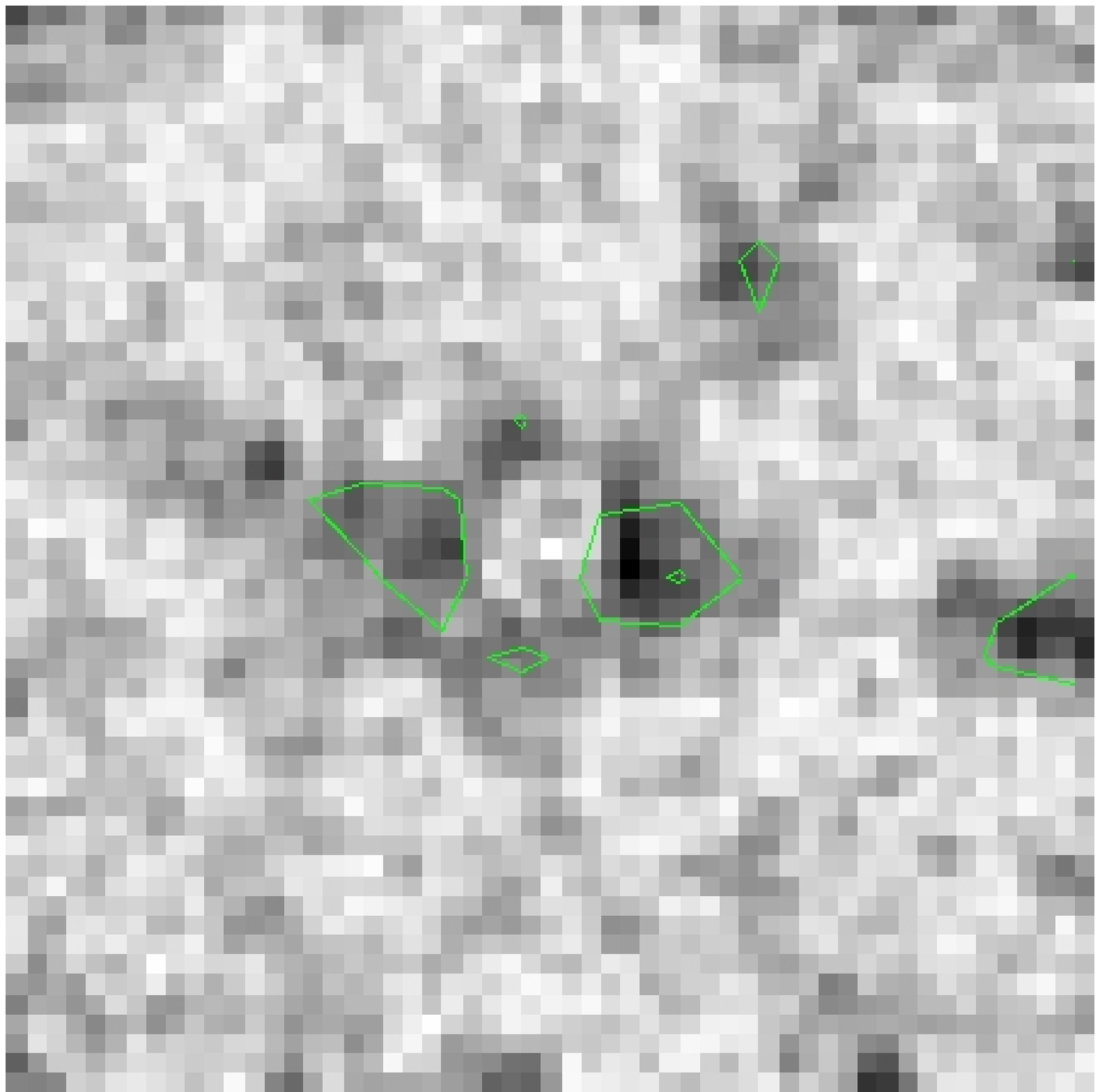}\\\\
\includegraphics[width=3.5in]{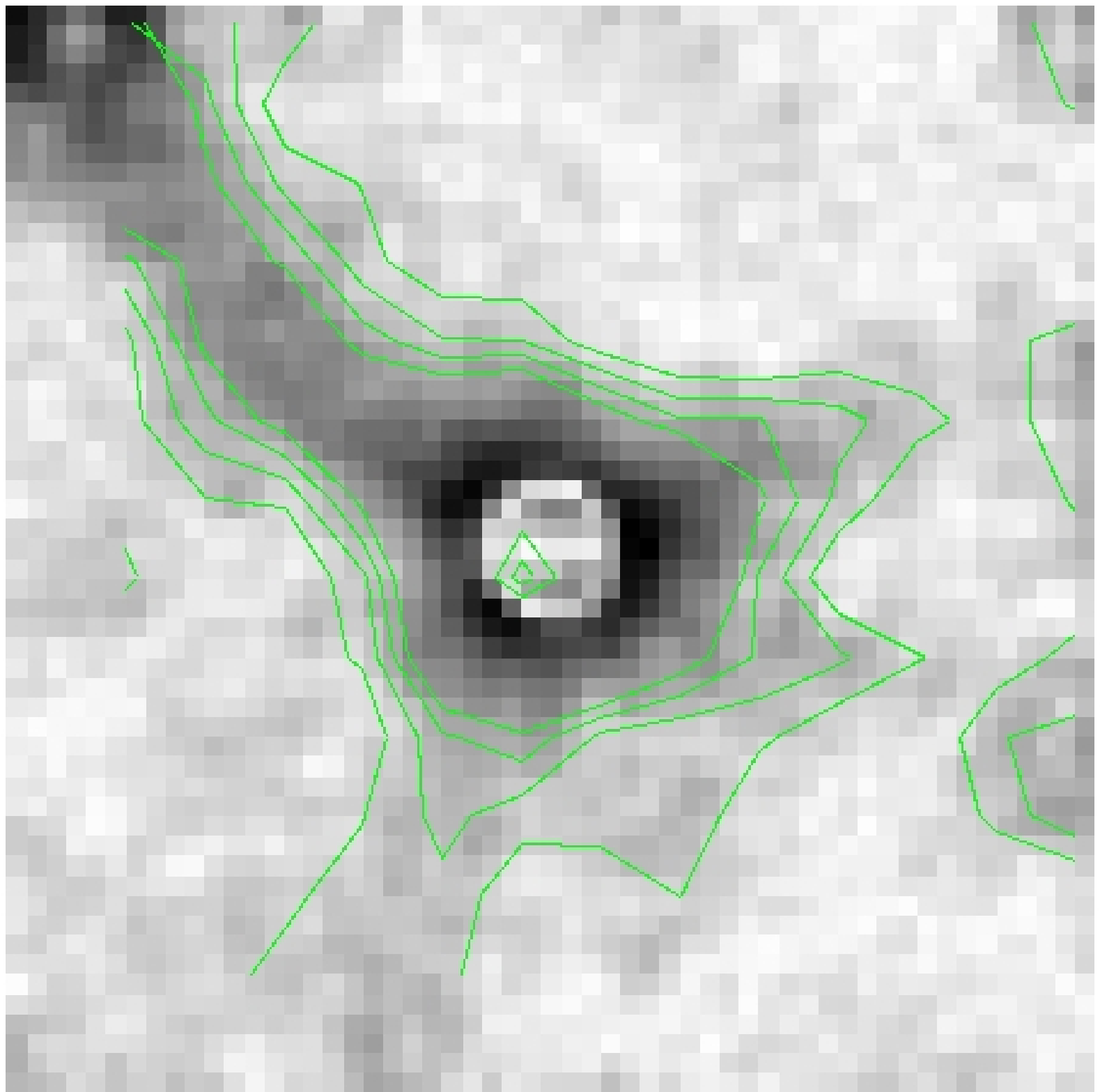}&
\includegraphics[width=3.5in]{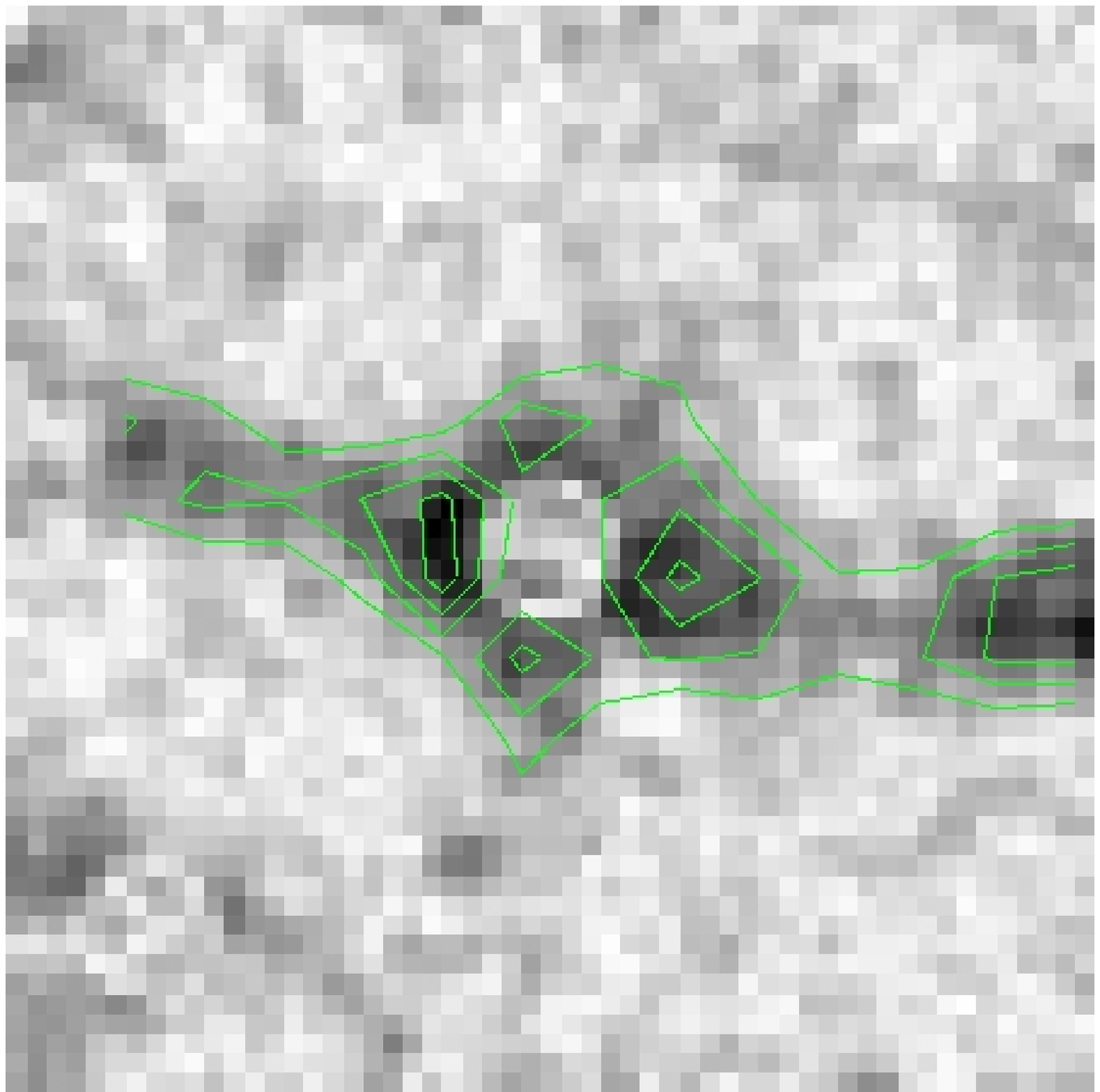}
 \end{tabular}
     \caption{Results for the  $n = 0$ shear filter using an aperture radius of 2.5\,$h^{-1}$ Mpc. The top panels show the cluster convergence map, on a logarithmic scale. The middle panels show a signal-to-noise map when the shear filter is applied to the cluster field with a lensed galaxy density of 30 arcmin$^{-2}$. The bottom panels show the same result, but this time for a galaxy density of 100 arcmin$^{-2}$. The contours correspond to signal-to-noise values of 3, 4 and 5. The fields are of side length 3.5$^{\circ}$, and the grid size upon which the filtering is performed is 100 $\times$ 100 pixels.}
     \label{fig:shearfilt}
\end{figure*}

The shear filter (eq.\,\ref{shearfilt}) was applied to a sample of clusters using the $n=0$ multipole and background galaxy densities of 30 arcmin$^{-2}$ and 100 arcmin$^{-2}$. The signal to noise maps corresponding to two cluster fields and an aperture radius of 2.5\,$h^{-1}$\,Mpc are presented in Fig.\,\ref{fig:shearfilt}. As before, the aperture mass was chosen to approximately correspond to the expected width of a filament. 

\begin{figure*}
\centering
\begin{tabular}{cccc}
          \includegraphics[width=2.7in]{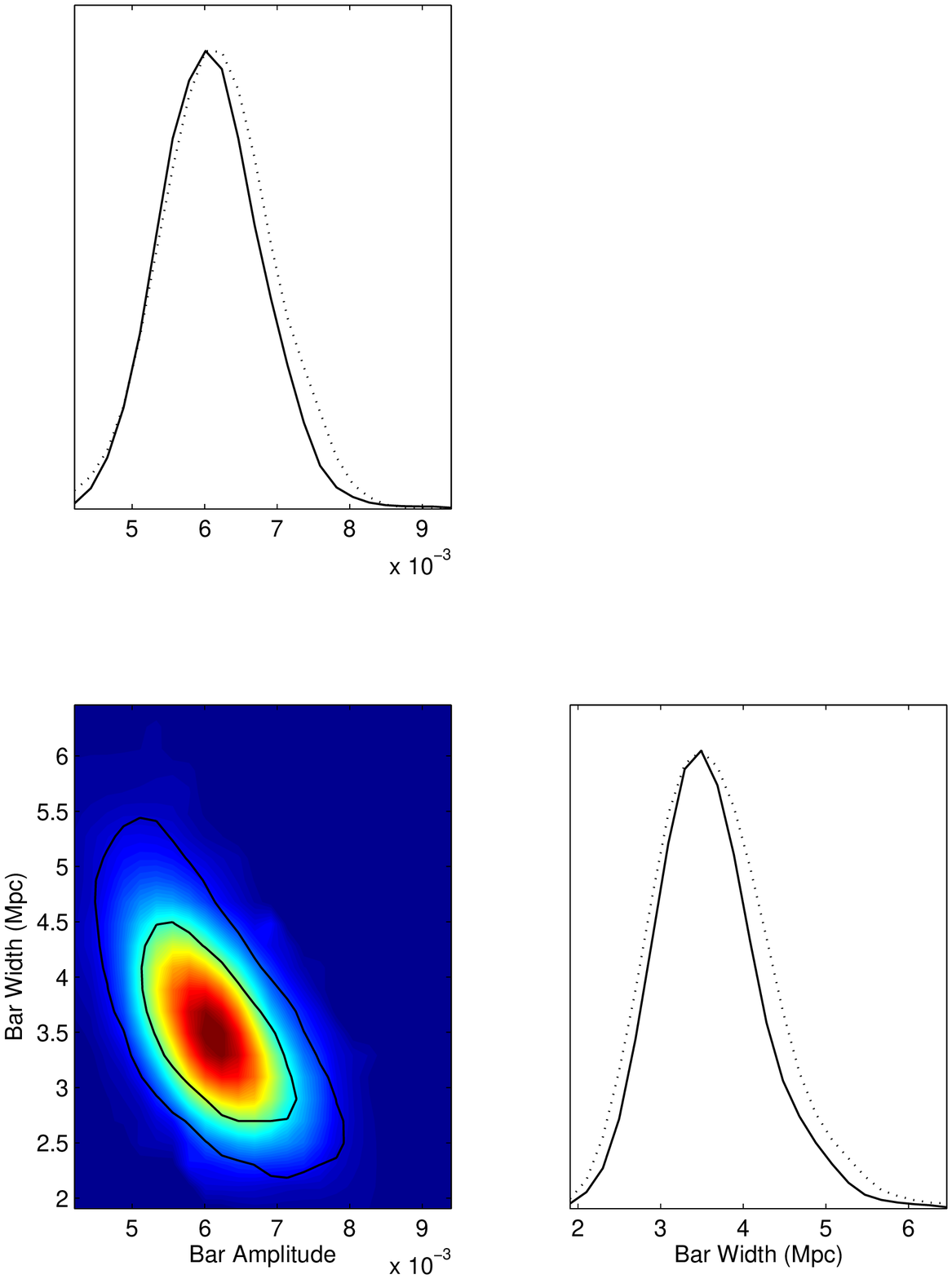}&&&
          \includegraphics[width=2.7in]{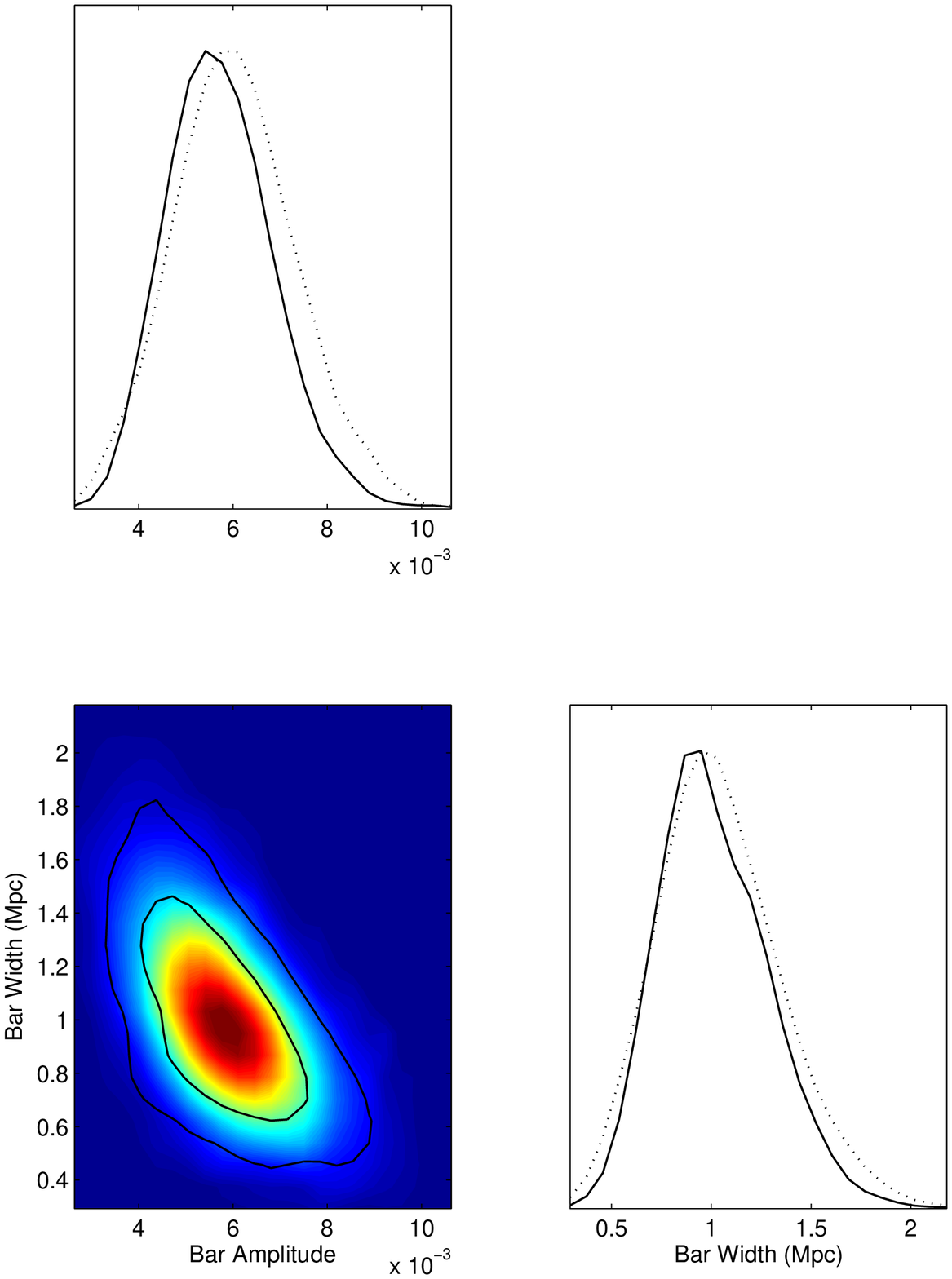}\\\\\\
           \includegraphics[width=2.7in]{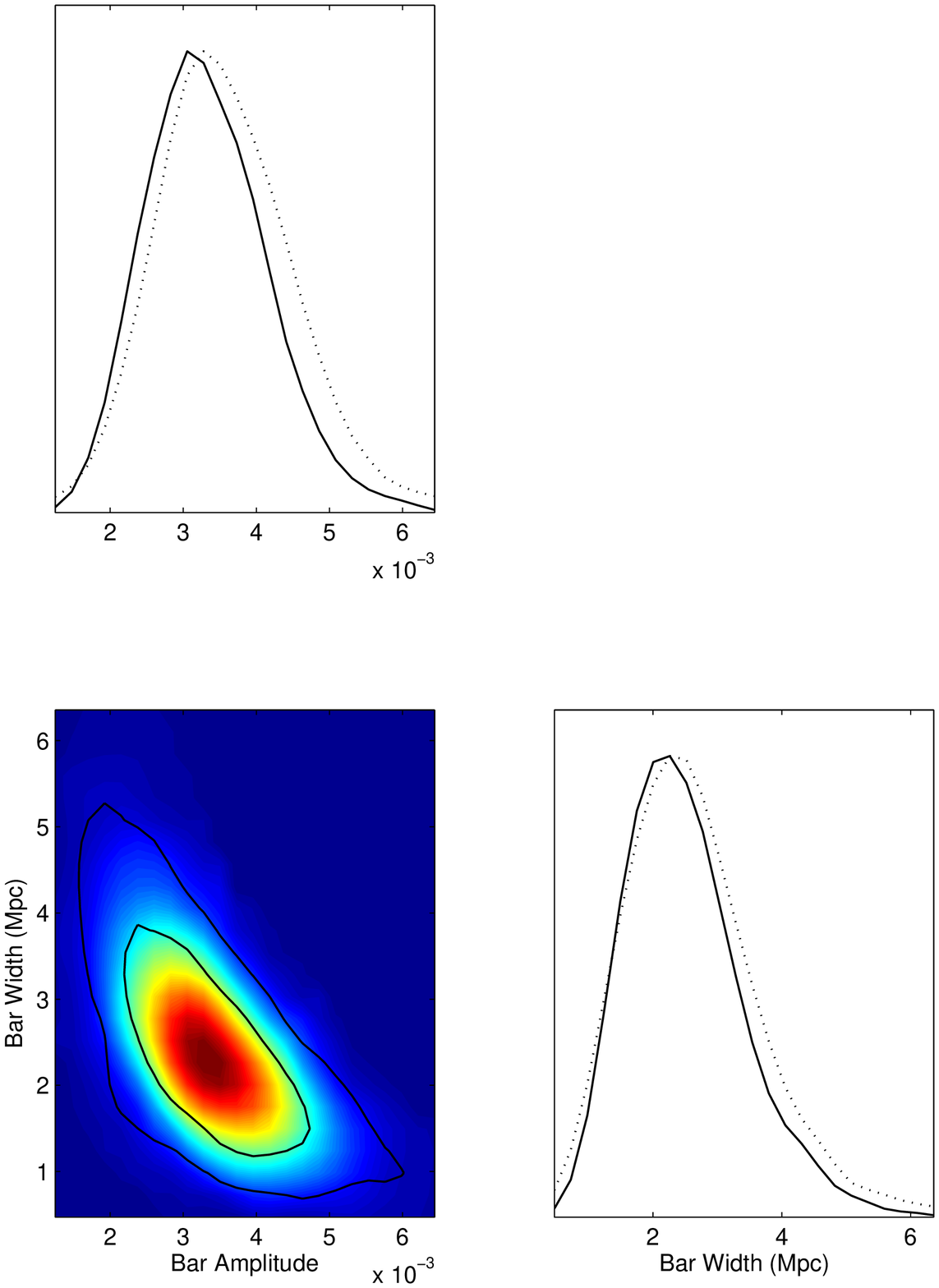}&&&
            \includegraphics[width=2.7in]{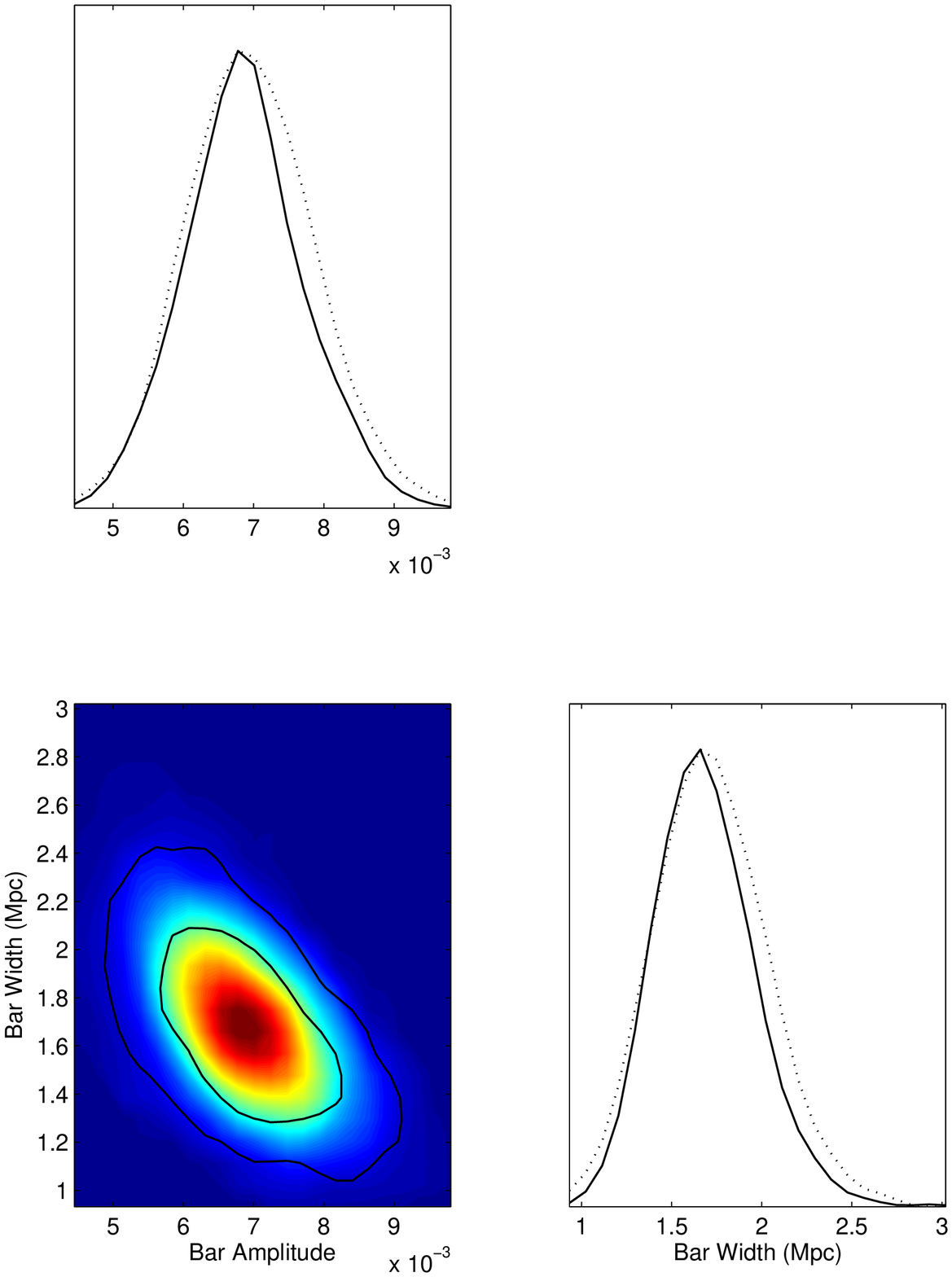}  
                \end{tabular}
     \caption{The four sets of panels show results of MCMC model fits to three different intercluster filaments: the 2D posterior PDF and 68 and $95\%$ confidence contours in bar amplitude and width (core radius), and their marginalized 1D distributions. All panels used a background galaxy number density of 100 arcmin$^{-2}$. }
\label{MCMCcontour}
\end{figure*}

The filtered results, in Fig.\,\ref{fig:shearfilt}, show clear evidence of filament detection. As expected, as galaxy density increases, filament detections are made at a higher signal-to-noise, and weaker filaments are more apparent. The strong filaments can be seen at both galaxy densities - at 30 arcmin$^{-2}$ the signal to noise along these strong filaments is approximately 3-5, and in certain regions of the filament even higher; at 100 arcmin$^{-2}$ the signal-to-noise along the length of these strong filaments is in the region 5-7, a very robust detection. Weaker filaments are less apparent at lower background galaxy densities, although in Fig.\,\ref{fig:shearfilt} there is evidence of weaker filaments in the 30 arcmin$^{-2}$ images. At 100 arcmin$^{-2}$ even these weak filaments come into view at signal-to-noise typically in the range 3-4 along the length of the filament.

The aperture size should be chosen to correspond roughly to the scale one wishes to investigate. Aperture sizes that are too small will produce weak signals, whereas if the aperture is too large it will pick up signal from mass unrelated to the structure of interest. In the specific case of detecting filaments, which are much weaker than the cluster,  any signal sufficiently close to the cluster should not be regarded as necessarily corresponding to filament detection, as such filters will enclose some of the cluster as well as any putative filament. To be confident of filament detection, the signal must at least extend outside $r_{200} + r_{\rm{max}}$, where $r_{200}$ is the virial radius of the cluster and $r_{\rm{max}}$ is the aperture size. In Fig. \ref{fig:shearfilt} we see that the asserted detections satisfy this criterion.

As discussed in earlier sections, using a single multipole is not the optimal method for detection. By superposing many multipoles, one can choose the coefficients in the expansion to maximize signal-to-noise. On investigation however, adding higher multipoles made only a small improvement to filament detection and is unnecessary.  One may also hope to improve the significance of a detection by altering the filter function, however, in this case attempting to match the filter function to expected filament profiles did not lead to a great improvement. The weight function in eq.\,(10) is suitable for filament detection.

\subsection{Application of MCMC Techniques}

We now consider the effectiveness of  MCMC techniques in detecting and describing filaments using a simple parameterized bar model, with a primary focus on those between cluster pairs. \cite{Colfil} identified filaments between neighbouring clusters in the N-body simulations of \cite{Kauffmann}. They found that  approximately $40\%$ of all filaments between clusters are straight and on centre with respect to the cluster-cluster axes; we shall term such filaments `regular'. Other filaments are `warped', irregular or off-centre. The former category of filaments are amenable to parameterized mass model fits. 

To select a sample of regular filaments, we selected pairs of clusters in our fields separated by a distance of less than $25\,{\rm h}^{-1} \rm{Mpc}$\footnote{ 
Note that in the $z\sim 0.2$ snapshot of the Millennium simulation, almost 25$\%$ of galaxy clusters with virial mass in excess of $2\times10^{14}{\rm h}^{-1} M_{\odot}$ (mass inside a radius where the mean density exceeds 200 times the critical density for the simulation) have a neighbour exceeding this mass limit.}.
Clusters such as this can be easily picked up on mass reconstructions, even at relatively low lensed background galaxy counts. Filaments connecting clusters tend to be straighter and more regular compared to those filaments which do not connect cluster pairs. A further advantage of focussing on filaments connecting `closely' spaced clusters is that since the clusters provide a start and end point to the filament, this means we can constrain to a good approximation both the filament's orientation and the position of its central axis. This reduces the number of parameters we have to fit in our mass model. 

As noted above we use {\small COSMOMC} (\cite{lewbri}) with a Metropolis algorithm 
sampler. We dynamically update the proposal density from the progressive covariance of 
post-burn-in samples, and we use standard convergence tests. Analysis of the chains produced uses {\small GetDist} distributed with {\small COSMOMC}.  Before fitting a mass model to the filament itself, NFW profiles were fit to the cluster halos, yielding values for their virial radii $r_{\rm vir}$.  We then used all galaxies which were a distance greater than two virial radii away from the cluster cores. This was to ensure that no artificial contribution to our bar fits was made by the clusters themselves.

\cite{Colfil} investigated the mass distributions of the filaments in their simulations, and determined that they tend to have cores of approximately constant surface density, after which their profiles are well approximated by a $r^{-2}$ decline. Therefore, we fit the profile,
\begin{equation}
\kappa(r) = \frac{\kappa_{0}}{1+\left(\frac{r}{r_{c}}\right)^{2}}\;,
\end{equation}
where $r$ is the perpendicular distance from the centre of the filament, and the amplitude of the model bar, $\kappa_{0}$, and the core radius, $r_{c}$, are the free parameters. The shear field corresponding to this bar can be derived using symmetry arguments. If we take one axis as lying along the filament (labelled `2') and the other axis orthogonal to the filament's length (labelled `1'), we know that the deflection potential cannot vary with respect to the former coordinate. This means we can set all derivatives with respect to this axis to zero in eq.\,(2). It is then clear that in this case,
\begin{equation}
\gamma_{1} = \kappa;
\quad \gamma_{2} = 0\;.
\end{equation}
We use flat priors on parameters, covering as broad a range as is physically possible, errors also being accounted for. In Fig.\,\ref{MCMCcontour} we show the likelihood contours obtained for three clusters, using a background galaxy density of 100 arcmin$^{-2}$. A fit to a `blank' field (a region of the simulation associated with neither the cluster nor a filament) was also performed, and as expected the MCMC fit gave core radii and bar amplitudes consistent with zero. \cite{Colfil} report that for straight filaments, the typical core radius of a filament lies between $1.5$ and $2.0\,h^{-1}$\,Mpc and our MCMC fits are consistent with this result.

The results were repeated for lower galaxy densities (30 arcmin$^{-2}$), and as expected, the general effect of this is to increase the spread in our parameter distributions. Although the fits were still inconsistent with an empty field, in the majority of cases the parameter range was so broad as to not provide physically meaningful constraints.

The filaments analyzed in Fig. \ref{MCMCcontour} all connect cluster pairs, but we may also apply this MCMC technique to filaments that do not connect closely spaced clusters. We took clusters which were not part of a cluster pair and applied the shear filter and mass reconstruction to the data fields, using a background galaxy population of 30 arcmin$^{-2}$. We then identified filament candidates from the results, and fit filament profiles to these candidates. We found that the resulting profiles gave core radii and amplitude in the expected range, with neither quantity consistent with zero.

\section{The Effects of Large Scale Structure}

As weak lensing is sensitive to all matter along the line of sight, the large scale structure (LSS) will make a contribution to any lensing signal. For instance, \cite{Hoekstra02} demonstrated that LSS is a major source of uncertainty in determining the virial mass and concentration parameters of clusters. This was confirmed by \cite{Dod04}, who also presented techniques to minimize the effects of the LSS. Given this previous work, we expect the presence of LSS to have a deleterious effect on the results presented in this paper. To investigate the influence of LSS on our results, we simulated Gaussian LSS fields with a  convergence power spectrum corresponding to the cosmology used in previous sections, and added them to our cluster simulations. The convergence range of these LSS fields are consistent with 
the findings of \cite{Whvale}, excluding regions where there are non-linear structures.  The inclusion of non-linear structures will thus increase the scatter beyond that of the analysis below. Additionally, large structures located along the line of sight could bias mass estimates. However, with photometric redshift estimates for sources in the field one could identify spikes in redshift space that are likely to impact on the results, and partially account for them in the models.

To quantify the effects of LSS on the shear filters, we can extend the results of \cite{s98}. In the presence of LSS, the ensemble average of the shear monopole will be unchanged. The dispersion due to LSS can be calculated as follows,

\begin{equation}
\ave{\left( \zeta^{(0)} \left(\bt \right) \right)^2}=\int \drm^2 \bt' U(\bt') \int \drm^2 \bt \ave{\gamma(\bt') \gamma(\bt)}\;,
\end{equation}

where the shear correlation function can be related to the power spectrum as described in e.g. \cite{SWM}, 

\begin{equation}
\ave{\gamma(\bf{0}) \gamma(\bt)} = \int_{0}^{\infty} \frac{\drm l \, l}{2 \pi} J_{4}(l \theta) \left[P_{E}(l) - P_{B}(l) + 2 \im P_{EB}(l) \right]\;.
\end{equation}

In this equation, $P_{E}$ and $P_{B}$ are the E- and B-mode power spectra outlined in \cite{SWM}. The dispersion due to different orders of multipole can be calculated in an analogous fashion.

When the shear filter was applied to the clusters with the addition of LSS noise, it was found that this noise confused and obscured filament detection. In particular, at a galaxy density of 30 arcmin$^{-2}$ filaments are difficult to distinguish from spurious signals due to the LSS noise - at this galaxy density we were unable to make any clear detections. The problems from the LSS are mitigated somewhat at higher galaxy densities, with strong filaments (i.e. those between cluster pairs) detected clearly at a galaxy density of 100 arcmin$^{-2}$.   

To investigate the effects of LSS on our MCMC analysis, we simulated a number of LSS noise realizations and used MCMC to find the best-fit parameter (the values for which there is a peak in the marginalized 1D distribution). It was found that the addition of LSS noise caused a spread in the values of the best-fit parameters, but there was no bias (i.e. the best-fit parameters were not systematically higher or lower). For instance, repeating an MCMC analysis of the cluster considered in the top left hand panel of Fig. 4  seven times we found that the width ranged between 2.8-5\,Mpc, and the amplitude ranged between 4.5-6.1 $\times 10^{-3}$. The results are still consistent with the presence of a bar; however, if the results are to be used to quantitatively analyze filament profiles the dispersion in the results needs to be reduced. In practice this could be done by statistically modeling the LSS structure noise and including it in the likelihood function.

\section{Discussion and Conclusions}

The typical strength of an intercluster filament is $\kappa \sim 0.01$, compared to a convergence in the central cluster regions of $\kappa \sim 0.5$. Therefore, the reason filaments are so difficult to detect using any method is due to the fact that they are much weaker than the cluster itself, and are quickly `lost' in the noise originating from the intrinsic ellipticity dispersion of the background galaxies. In this paper, we have shown several methods that can be used to reliably detect filaments. By implementing a variety of methods, we have also shown how versatile weak gravitational lensing is as a tool for detecting filaments - each method described in the paper has different strengths and weaknesses that make them suited to different data sets. 

The mass reconstruction is one of the most common methods to detect overdensities in a mass distribution. We applied the finite field mass reconstruction algorithm of \cite{seitzrec} to both single and stacked clusters. For single clusters, reconstructions using a background galaxy density of 30 arcmin$^{-2}$ weakly detect only the strongest filaments, with the majority of filaments undetected. As background galaxy density increases, so does the ease of filament detection - at 100 arcmin$^{-2}$ we are able to identify some filaments in reconstructions. The mass reconstruction is most effective when used with stacked data sets. The stacking method has an impact on the quality of the results, with the double-cluster method proving more effective than aligning cluster major axes. A major disadvantage of mass reconstructions is that they can give false filamentary structures due to the smearing effect of the smoothing scale. This is not a problem unique to mass reconstructions, and any method which smoothes results over certain scales in an aperture will suffer from the same difficulty. This is a particular problem for closely-spaced clusters, which will appear to merge in the reconstructed image if the smoothing scale is greater than the inter-cluster separation. The second major disadvantage of a mass reconstruction is that it does not allow easy calculation of the significance of any filament detection. The origin of this effect lies in the fact that in any reconstruction, the shear field must be smoothed to avoid infinite noise \citep{ks93}. This leads to correlated errors in the resulting convergence map, making error bars very difficult to construct. Thirdly, because filaments are weak, they are often located in regions that are dominated by noise. As \cite{Dietrich} points out, in such a scenario the value of the mass sheet degeneracy can fluctuate by as much as the filament strength between different galaxy populations, making any filament detection in an individual cluster more questionable. This third point does not apply for the stacked clusters, as the stacking ensures that noise is reduced to sub-dominant levels even in filamentary regions.  Therefore, it is advisable to use the mass reconstruction in conjunction with other methods described in this paper, both to confirm any detection and to allow the significance of any detection to be accurately assessed. 

In general, the multipole moments of the convergence peformed poorly, and were unable to register any strong filament detections in single clusters. Although \cite{Dietrich} used $Q^{(2)}$ to provide evidence for a filament in A222/223, this cluster pair was very closely spaced and the corresponding filament was thus much stronger than any considered in this paper. It is also unclear how much of a contribution the filament makes to the quadrupole signal, as it is likely that the majority of the signal comes from the clusters themselves. The quadrupole signal fails to detect filaments in the stacked data due to misalignments. Misalignment between filaments will create a `mass-sheet' in the resulting stacked data, which will give zero quadrupole signal. 

It is not only possible to define multipole moments of the convergence, but also of the shear. Here we have shown that the shear monopole is very effective in picking out filaments. The reason for this is that the shear monopole performs a (weighted) sum of all shears in an aperture. If these shears are coherent over the aperture, then the net result will be a large signal. The shear signals corresponding to a filament will all point in approximately the same direction along the filament's length, and thus filaments are well suited to detection using this method. However, this method is not tailored to exclusively find filaments, and a variety of mass distributions will give a strong signal. The LSS adds noise to the shear monopole signal which can, in the cases of weak filaments or low galaxy densities, cause confusion, false-identification and obscuration of the filament signal. However, we have shown that at galaxy densities characteristic of future space-based missions, or if particularly strong filaments are targeted, the filament signal will still stand out clearly against the LSS noise.

We demonstrated that one can use MCMC techniques to fit filament profiles, even at relatively low galaxy densities. This is an important result as it shows that weak lensing can be used not only for filament detection, but also for quantitative analysis of filament properties. In the future, this will allow us to directly compare the predictions of simulations to observations. The scatter in the best fit parameters caused by LSS has been quantified, and hinders the ability of this method to quantify profiles accurately - in future work, the LSS should be incorporated into the likelihood function to minimize the impact on the results (e.g. as per the prescription laid out in \citet{HuWh}). The MCMC method was also shown to be quite versatile. It can be used on its own, as it was when we looked for filaments between cluster pairs, or it can be used in conjunction with other techniques such as the mass reconstruction or in particular the shear filter to first of all search for a filament candidate. There are a number of caveats associated with using MCMC fits. Firstly, it is model-dependent - this involves assuming a profile for the filament, in this paper based on the findings of \cite{Colfil}. Models that more accurately reflect the `true' filament profiles will achieve better likelihood values. Secondly, since it is very difficult to write down a model for irregular or warped filaments, this method is only suitable for investigating the properties of straight, regular filaments. This represents only a fraction of the total filament population. Thirdly, the method works best if we have some prior information on the location of the filament, either from one of the other detection methods, or by using cluster pairs between which filaments are thought to be constrained. Without a prior on the filament position, fitting any profile would become a complicated procedure.

Stacking is a technique that can be used in conjunction with any of the methods described above, to improve the chance of filament detection. Stacking is particularly useful when each individual cluster lenses a low background galaxy density, but the survey contains many such clusters. In this paper we suggested two possible stacking methods: the first aligned cluster major axes in the stacked images and the second aligned axes joining cluster pairs. The latter method is far more effective. This is because, although filaments are often observed to align with the major axis of a cluster, there is no guarantee that they are either regular or strong. In almost all cases there is imperfect alignment of the filaments with the cluster major axis, and sometimes there is no filament aligned in this direction at all. If irregular or misaligned filaments are stacked the net effect will be to produce a low density mass sheet. This is not ideal for any of the methods described in this paper. On the other hand, filaments between cluster pairs tend to be straight, strong and regular and hence are easily stacked. The only disadvantage of using cluster pairs is their relative scarcity - whereas any cluster can be used when aligning the major axes, fewer cluster pairs will exist in a survey. 

In summary, in this paper we have presented a variety of new detection methods, some of which (for example the shear filter and MCMC) give promising results. However, due to the low surface densities of filaments, and the additional effects of LSS noise, the task of detecting filaments in actual data remains difficult. In reality, filament detection will either require the high galaxy densities of future space-based missions, or the stacking of a number of clusters. Current space-based lensing studies achieve $\approx$ 80 galaxies arcmin$^{-2}$ (e.g. \citep{Leonard}), and we could achieve in excess of 250 arcmin$^{-2}$ for deep space-based observations (e.g. \citep{Rhodes}). Although there is a trade-off between the area and depth of a survey, so that those targeting many thousands of square degrees will typically not go so deep, the most massive clusters such as those considered in this work will be prime targets for deep targeted observations. Stacking the data from the fields of less massive clusters in non-targeted large area surveys will also be extremely useful in statistically constraining the properties of filaments. 
 
There is one further method, not discussed in this paper, that could prove valuable in detecting filaments -  gravitational flexion. In traditional weak lensing studies, such as the formalism used in this paper, the lens equation is approximated as linear. This means that any distortion in the ellipticity of a galaxy will be aligned purely tangentially to the lens. Physically, such a linear approximation is equivalent to assuming no variation in the lens field over the scale of the lensed image. However, if we do not make such an assumption, the lens equation becomes non-linear, and in addition to the convergence and shear, gravitational distortion is determined by the first and second flexion. Taken together, flexion introduces `arciness' to the image and some radial alignment with respect to the lens (see \cite{GB05} for further details). It has been demonstrated, for example by \cite{LKW09}, that gravitational flexion is an effective probe of substructure in galaxy clusters. This suggests that flexion may also be a useful tool for filament detection. 

Besides detecting filaments using weak lensing data, as we have noted in the Introduction, there are 
also a variety of other complementary detection methods. Sunyayev-Zeldovich observations could in future be used in combination with more established methods of filament detection (e.g. weak lensing, X-ray, galaxy overdensity) to provide more optimal algorithms for detection and even stronger constraints on profiles. The area of overlap between DES and the South Pole Telescope, SPT, will result in several tens of thousands of galaxy clusters in common. The ability to probe the total matter content of clusters and their environments, coupled with deep optical and infrared imaging and spectroscopic observations, will enable us to determine galaxy bias as a function of environment. It will also improve our understanding of the formation of the most massive bound objects in the universe. Weak lensing of background galaxy populations becomes less effective for use as a probe of higher redshift structures \citep{lewking} and beyond $z\sim 1$ an alternative source population will be essential -- such as the 21cm emission from high redshift proto-structures that will be imaged with SKA, or the CMB as seen by a future mission. The ability to trace the evolution of filaments in the cosmic web will have important consequences for models of structure formation e.g. how much material is bound in filaments as a function of epoch.

\section{acknowledgments}
We would like to acknowledge the anonymous referee for helpful suggestions that have improved this paper. We'd also like to thank Antony Lewis, Damien Quinn, Tom Theuns and Patrick Simon for many helpful discussions. JMGM thanks STFC for a postgraduate award. LJK and IGM thank the Royal Society and the Kavli foundation respectively. The Millennium Simulation data was obtained from the Institute for Computational Cosmology at Durham University. The authors thank Carlos Frenk, Simon White, and Volker Springel for  allowing them access to the data and John Helly for his assistance with retrieving it.

\appendix

%
\label{lastpage}

\end{document}